\documentclass[a4paper,12pt]{article}
\usepackage{authblk}
\usepackage[square,sort,comma,numbers]{natbib}
\usepackage{soul}
\usepackage{color}
\usepackage{pifont}
\usepackage{mathptmx}
\usepackage{graphicx}
\usepackage{amsmath, amsfonts, amssymb, mathrsfs}
\usepackage{ccaption}
\usepackage{fixmath}
\usepackage{tabu}
\usepackage{longtable}

\DeclareMathOperator*{\argmin}{arg\,min}
\usepackage{graphicx}
\DeclareMathOperator{\logit}{logit}
\usepackage{multirow}
\usepackage{multicol}
\usepackage[figuresright]{rotating}
\usepackage[framemethod=tikz]{mdframed}
\usepackage{xcolor}
\usepackage{bm}

\title{Effects Among the Affected}
\author[1,2]{Lina M. Montoya \thanks{Email address for correspondence: lmontoya@unc.edu}}
\author[3]{Elvin H. Geng}
\author[2]{Michael Valancius}
\author[2,4]{Michael R. Kosorok}
\author[5]{Maya L. Petersen}
\affil[1]{School of Data Science and Society, University of North Carolina at Chapel Hill, U.S.A.}
\affil[2]{Department of Biostatistics, University of North Carolina at Chapel Hill, U.S.A.}
\affil[3]{Division of Infectious Diseases, Washington University in St. Louis, U.S.A.}
\affil[4]{Department of Statistics and Operations Research, University of North Carolina at Chapel Hill, U.S.A.}
\affil[5]{Division of Biostatistics, University of California, Berkeley, U.S.A.}

\date{August 2024}

\begin{document}

\maketitle
\pagebreak

\begin{abstract}
Many interventions are both beneficial to initiate and harmful to stop. Traditionally, to determine whether to deploy that intervention in a time-limited way depends on if, on average, the increase in the benefits of starting it outweigh the increase in the harms of stopping it. We propose a novel causal estimand that provides a more nuanced understanding of the effects of such treatments, particularly, how response to an earlier treatment (e.g., treatment initiation) modifies the effect of a later treatment (e.g., treatment discontinuation), thus learning if there are effects among the (un)affected. Specifically, we consider a marginal structural working model summarizing how the average effect of a later treatment varies as a function of the (estimated) conditional average effect of an earlier treatment. We allow for estimation of this conditional average treatment effect using machine learning, such that the causal estimand is a data-adaptive parameter. We show how a sequentially randomized design can be used to identify this causal estimand, and we describe a targeted maximum likelihood estimator for the resulting statistical estimand, with influence curve-based inference. Throughout, we use the “Adaptive Strategies for Preventing and Treating Lapses of Retention in HIV Care” trial (NCT02338739) as an illustrative example, showing that discontinuation of conditional cash transfers for HIV care adherence was most harmful among those who most had an increase in benefits from them initially.
\end{abstract}

\maketitle

\section{Introduction}

Many treatments appear to be beneficial to receive \emph{and} harmful to stop -- yet they must be delivered in a time-limited way. For example, long-term opioid use for chronic pain management can lead to dependence and tolerance. Even with proper tapering, patients who initially benefited may experience withdrawal symptoms and increased pain sensitivity, making pain management more challenging post-discontinuation \citep{kuntz2021factors}. Programs that promote healthy behaviors in workplace settings may lose their effectiveness once the program ends, as some employees initially helped return to their old habits if the motivation was primarily driven by program incentives \citep{cancelliere2011workplace}. Nutritional support initiatives can improve dietary habits, but discontinuation may lead to a return to previous, less healthy eating habits, especially in segments of the population that depend on these programs for access to healthier food options \citep{atoloye2021effectiveness}.

Whether or not to deem a time-limited intervention worthwhile for a population traditionally comes down to the simple of question of whether, on average, it provides a net benefit; in other words, whether the expected benefit of starting outweighs the expected harm of stopping. However, such an apparently straightforward finding risks missing nuances that can come into focus once examined through the lens of precision medicine \citep{kosorok2019} and, more generally, treatment effect heterogeneity -- specifically, if the harms of discontinuing an intervention may occur in different (or the same) people as those for whom the intervention is initially helpful. Consider a net beneficial intervention. It may be the case that initiating the intervention has no effect on some but helps others; and, once removed, the harms of discontinuation only or primarily accrue among those who initially benefited from the intervention. In other words, the time-limited intervention may provide a temporary benefit for some, a sustained benefit to others, but harms no one. On the other hand, it may be the case that the harm of treatment discontinuation accrues in people who were not benefited to begin with; i.e., the persons harmed by treatment removal are distinct from those helped by starting.

While from a purely utilitarian perspective such correlations are irrelevant, identifying and distinguishing between these two scenarios, and more generally quantifying the extent to which the harms of stopping an intervention are correlated with the benefits of starting it, is important both for understanding the equity impacts of deploying an intervention and for understanding its mechanism of action.  An intervention that provides a time-limited benefit to some, and perhaps a sustained benefit to a smaller number, could be deployed to a full population (cost and resource constraints aside) without risking harm to any. In contrast, an intervention whose removal harms those not benefited to begin with requires additional tailoring before deployment to avoid doing harm.

Longitudinal data can allow us to quantify the extent to which the benefit of treatment initiation modifies the effect of subsequent  treatment discontinuation. In particular, Sequential, Multiple Assignment Randomized Trials (SMARTs) permit asking such causal questions because of the conditional randomization that occurs at several stages. Often, a SMART's sequential conditional randomization is leveraged to identify treatment effect heterogeneity, by baseline and time-varying covariates, that informs point-treatment or longitudinal (optimal) dynamic treatment regimes (e.g., see \cite{kosorok2015adaptive, murphy2005experimental, kidwell2023sequential}). However, in this paper, we suggest that SMARTs are also particularly well-suited to answer questions about the heterogeneous distribution of the harms and benefits from time-limited treatments. 

As an illustrative example, consider the SMART called Adaptive Strategies for Preventing and Treating Lapses of Retention in HIV Care (ADAPT-R) trial (NCT02338739; \cite{geng2023adaptive}), carried out to ultimately identify strategies for helping patients remain retained in their HIV care. In this study, adults living with HIV in rural Kenya were sequentially randomized at two time-points to receive incentives that aimed to reduce the occurrence and/or recurrence of a lapse in their care. Specifically, at each stage, patients were randomized to receive either a conditional cash transfer (CCT) for on-time clinic appointments or a non-CCT based intervention (such as SMS text messages, peer navigators, or standard-of-care treatment). Under this design, it is straightforward to identify heterogeneous treatment effects of, for example, discontinuing CCTs at the second time-point after receiving them at the first time-point by baseline (e.g., age, sex, SES) and/or time-varying (e.g., viral load status, updated pregnancy status) variables. Indeed, there exist a range of machine learning algorithms to estimate such heterogeneous treatment effects (see, e.g., \cite{wager2018cf, semenova2021debiased, kennedy2023towards}).

In this paper, we explore this type of downstream, second-stage effect modification not by a static set of variables, but rather by a function of initial treatment response -- namely, the conditional average treatment effect (CATE). This allows us to ask our original question: how do the effects of discontinuation differ between those for whom treatment initiation was harmful versus helpful? Or if treatment was helpful for all, how does the effect of discontinuation differ between those with large versus attenuated benefits? For example, in ADAPT-R, it was found that CCTs were beneficial on average -- marginally and conditional on baseline covariates. Further, it was found that, among people who did not have an initial lapse in care, continuing CCTs was beneficial, on average, compared to discontinuing CCTs. Here, we are interested in uncovering differential effects of continuing versus discontinuing CCTs as a function of participants' initial benefit from CCT initiation. In this way, we are able to investigate incentive discontinuation harms among those for whom the likelihood of initial benefits are large versus small. It may be the case, for example, that having an initial strong effect of CCTs leads to greater ``withdrawal" of CCTs -- which may be expected if CCTs are operating primarily via directly addressing structural barriers present only for a subset of the population. Or it may be the opposite: having initial strong effects of CCTs may be protective against the effects of their removal, which could occur if the CCTs' primary mechanism was via increasing intrinsic motivation for different segments of the population. Beyond mechanistic insights, a finding of maximal harm among persons with minimal initial benefit would highlight the risks of deploying a time-limited CCT intervention at scale to a subset of the population (without additional tailoring at either the initiation or discontinuation stages).

To this end, we introduce a causal parameter that helps us understand differential downstream consequences of initial, upstream treatment effects. In particular, we consider a marginal structural working model \citep{robins1999association, neugebauer2007nonparametric} that is a function of an estimated CATE, and define a data-adaptive parameter of interest \citep{hubbard2016} defined by this working model. Following the causal roadmap \citep{petersen2014causal}, we then identify this parameter as a function of the observed data distribution, and present an estimation procedure for our target parameter using Targeted Maximum Likelihood Estimation (TMLE; \cite{petersen2014targeted, rosenblum2010targeted}), with influence-curve based inference on estimates. We note that while presented in the context of a time-limited treatment, the causal estimand, identification approach, and estimator proposed to study effect modification of future treatment effects by prior response extends to the more general case of static sequenced interventions with the potential for both benefit and harm. In short, this type of analysis can allow us to examine effects among the (un)affected.

The rest of this paper is organized as follows. In Section 2, we describe the data generated from a SMART and the causal model that generates this data, with a description of the ADAPT-R data, which motivates the development of these methods. In Section 3, we define our causal question of interest (are there effects among the [un]affected?), as well as the target causal parameter that corresponds to the specified causal question of interest (the parameters of a marginal structural working model) that summarizes how the effect of treatment discontinuation varies as a function of initial treatment response. In particular, we use machine learning to estimate the conditional average treatment effect (CATE) for treatment initiation, and the estimated CATE as a baseline effect modifier, resulting in a data-adaptive target parameter. In Section 4, we discuss the conditions needed for identification of the causal parameter of interest, and how data generated from a SMART can facilitate this identification process. In Section 5, we discuss a TMLE of the parameters of the identified marginal structural model, as well as influence-curve based inference for this estimator. In section 6, we present two simulation studies illustrating performance of this estimator in finite samples. In Section 7, we apply these methods to the ADAPT-R study. We close with a discussion.

\section{Data \& Causal Model}

The following data are generated from a $K$-stage SMART for a time $t \in \{0,..K\}$, where $K$ corresponds to the number of randomization stages: 1) categorical interventions $A(t) \in \mathcal{A}_t$ (which could include missingness or measurement variables, such as a time-varying indicator of right-censoring); and, 2) non-intervention variables $L(t) \in \mathcal{L}_t$ between interventions at time $t-1$ and $t$, which include baseline covariates $L(0)$, time-varying covariates $L(t)$, $t \neq \{0,K\}$ (which includes intermediate outcomes $Y(t) \in \mathbb{R}, t \neq \{0,K\}$), and a final outcome $L(K) = Y(K)$. Here, $L(-1) = A(-1) = A(0) = \emptyset$ and overbars are used to denote a variable's past history, i.e., $\bar{A}(t) = (A(1), \ldots, A(t))$ and $\bar{L}(t) = (L(0), \ldots, L(t))$. 

The subset of observed variables that are used to assign treatment at time $t$ according to the SMART's design are denoted $\bar{Z}(t) \subseteq (\bar{A}(t-1), \bar{L}(t-1))$. With this, the following structural causal model (SCM, denoted $\mathcal{M}^F$) describes the process that gives rise to data generated from a SMART design \citep{pearl2000}, where the random variables in $\mathcal{M}^F$ follow the joint distribution $P_{U,X} \in \mathcal{M}^F$:
\begin{align*}
    L(t) &= f_{L(t)}(U_{L(t)}, \bar{L}(t-1), \bar{A}(t)) \\
    A(t) &= f_{A(t)}(U_{A(t)}, \bar{Z}(t)),
\end{align*}
for $t = 0, ..., K$. Here, $U = (U_{L(t)}, U_{A(t)})$ represents the unmeasured random input to the data generating system at time $t$. In a SMART, the functions $f_{A(t)}$ are known for all $t$, as they are the randomization scheme used in the SMART. Further, $U_{A(t)}$ is known by design in a SMART to be independent of all other unobserved error terms. 

\subsection{ADAPT-R Data}\label{adapt-data}

We refer the reader to \citep{geng2023adaptive, montoya2023efficient} for details on the ADAPT-R study. Briefly, adults living with HIV in Kisumu, Kenya were initially randomized with equal probability to interventions intended to prevent lapses in HIV care (short message service [SMS] messages, conditional cash transfers [CCTs], or standard-of-care [SOC] counseling). If patients had a lapse in HIV care, they were re-randomized with equal probability to a more intensive intervention (SMS and CCTs, peer navigator, or SOC outreach), intended to re-engage them back into care. If patients succeeded in their initial care and were initially randomized to an active arm, they were re-randomized (with equal probability) to either continue or discontinue the initial intervention. Those who succeeded and were initially given SOC remained in SOC. 

\begin{figure}[h]
    \centering
    \includegraphics[scale = .4]{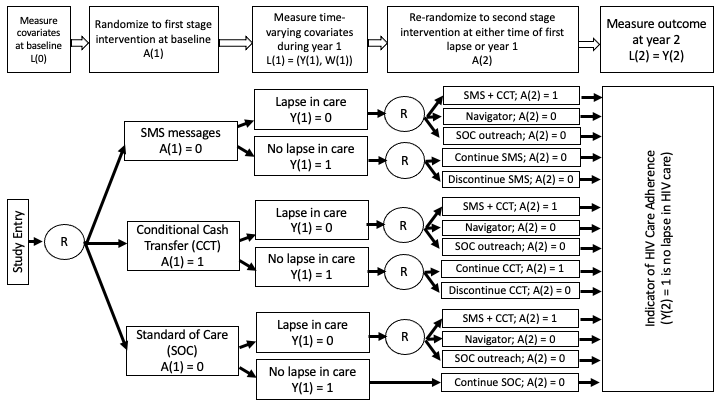}
    \caption{Adaptive Strategies for Preventing and Treating Lapses of Retention in HIV Care (ADAPT-R) trial design.}
    \label{p0}
\end{figure}

Letting $t = 1$ be the time of first randomization and $t = 2$ be the time of second randomization (either date of first retention lapse or one year after initial randomization, whichever occurs first), the observed ADAPT-R data consisted of the following:
\begin{itemize}
    \item Baseline covariates $L(0)$ (the full list of baseline covariates are in Appendix A);
    \item Stage 1 prevention intervention $A(1)$, which consisted of either receiving CCTs (coded as $A(1)=1$), or not receiving CCTs (i.e., receiving SOC or SMS; coded as $A(1)=0$);
    \item Time-varying covariates assessed between randomization to Stage 1 and Stage 2 interventions, $L(1) = (Y(1), W(1))$, which included: 1) $Y(1)$, an indicator of whether there was a lapse in care ($\geq$ 14 days late to a clinic visit) within the first year after enrollment, where $Y(1) = 1$ is no lapse in care. $Y(1)$ in this case is also the first-stage outcome; 2) $W(1)$, other time-varying covariates unrelated to the randomization scheme, such as time from first randomization to second randomization;
    \item Stage 2 retention intervention $A(2)$, which consisted of either a CCT-based intervention (coded as $A(2) = 1$) or a non-CCT based intervention (coded as $A(2)=0$);
    \item The outcome of interest $L(2) = Y(2)$, an indicator of no lapse in care (i.e., no visits made $\geq 14$ days late) within the year following re-randomization to second-stage treatment, where $Y(2) = 1$ is no lapse in care.
\end{itemize}  
A schematic of the study and how the data were generated is presented in Figure \ref{p0}. The SCM relating these variables is covered by the general SCM above, for $t = 2$.

\section{Causal Question and Parameters}

The causal question of interest asks: do the initial effects of an earlier intervention modify the effects of a subsequent intervention? Concretely, for some $r < s$, is the average effect of $A(s)$ on $Y(s)$ moderated by the average effect of $A(r)$ on $Y(r)$ conditional on patient history prior and up to time $r$, i.e., $H(r) \equiv (\bar{L}(r-1), \bar{A}(r-1))$?  

In the context of ADAPT-R, our scientific question is the following causal question: does the initial, individualized effectiveness of initial CCT modify its discontinuation effects? Is it the case that incentive discontinuation (i.e., stopping CCT administration) harms those for whom the probability of initial benefit is small versus large? In this section, we will translate this causal question into a causal parameter $\Psi^{F}(P_{U,X})$, where $\Psi^F$ is a mapping from the causal model to the parameter space $\mathbb{R}$.

\subsection{Defining, Identifying, and Estimating the Conditional Average Treatment Effect}\label{identification_cate}

To begin to answer these questions, we first intervene on the SCM to generate individualized, initial effects of interest at time $r$; for $t = 0,...,r-1, r+1,..., K$:
\begin{align*}
    L(t) &= f_{L(t)}(U_{L(t)}, \bar{L}(t-1), \bar{A}(t))\\
    A(t) &= f_{A(t)}(U_{A(t)}, \bar{Z}(t))\\
    A(r) &= a(r) \in \{0,1\} \\
    L(r) &= f_{L(r)}(U_{L(r)}, \bar{L}(r-1), \bar{A}(r)),
\end{align*}
where the support of $A(r)$, i.e., $\mathcal{A}_r$, must be $\{0,1\}$. In this way, we can define the CATE at time $r$ for a counterfactual outcome $Y(r)_{a(r)}$ in $L(r)$ as: 
\begin{align}
    B_{P_{U,X}}(H(r)) &\equiv \mathbb{E}_{P_{U,X}}[Y(r)_{a(r)=1} - Y(r)_{a(r)=0} | H(r)], \label{blip_causal}
\end{align}
where $Y(r)_{a(r)}$ is the outcome at time $r$ if, possibly contrary to fact, an individual received treatment $a(r)$. 

Let $P_0$ be the true distribution of our observed data $O$, an element of $\mathcal{M}$, the statistical model. We assume the observed data $O_i \equiv (\bar{L}(K)_i, \bar{A}(K)_i) \sim P_0 \in \mathcal{M}$, $i = 1, \ldots, n$ were generated by sampling $n$ i.i.d. copies from a data-generating system contained in $\mathcal{M}^F$ above.

In a SMART, the randomization condition $Y(t)_{a(t)} \perp A(t) | \bar{Z}(t)$ holds for $t = 1,\ldots, K$, because $A(t)$ is conditionally randomized based on values of $\bar{Z}(t)$ (which may be $\emptyset$, in which case $A(t)$ is marginally randomized). Informally, this means there are no unmeasured common causes between $Y(t)$ and $A(t)$ at each timepoint, conditional on $\bar{Z}(t)$. We can extend this to state $Y(r)_{a(r)} \perp A(r) | H(r)$.

According to the SMART design, at certain $\bar{Z}(t) = \bar{z}(t)$ it is the case that $g_{A(t),0}(A(t)=a(t)|\bar{Z}(t) = \bar{z}(t)) > 0 - a.e.$ holds for each $t = 1,\ldots, K$ and $a(t) \in \mathcal{A}_t$, where $g_{A(t),0}(A(t)|\bar{Z}(t) = \bar{z}(t)) = P_0(A(t)|\bar{Z}(t) = \bar{z}(t))$, the true conditional probability of the treatment at time $t$ given accrued treatment and covariate variables that define the SMART's design. We consider CATEs at time $r$ for which the relevant positivity condition is not violated, in which case we can say $g_{A(r),0}(A(r)=a(r)|H(r)) > 0 - a.e.$ for $a(r) \in \{0,1\}$. Informally, this positivity condition states that, conditional on patient history (including randomization triggers), there must be a positive probability of receiving treatment $A(t) = a(t)$. It is important to consider this positivity condition to understand which CATEs at time $r$, i.e., $B_{P_{U,X}}(H(r))$, are feasible to examine. For example, in ADAPT-R, the CATE of $A(1) \in \{\text{CCT, no CCT}\}$ on $Y(1)$ is identifiable because $A(1)$ was marginally randomized and thus all patients could have theoretically received all possible treatments in $A(1)$. 

Both the randomization and positivity conditions permit identifying the initial CATE function (i.e., equation \ref{blip_causal}) as the so-called ``blip" function: $B_{P_{U,X}}(H(r)) = B_0(H(r))$. Let $B_{n}(H(r))$ be an estimate of $B_{0}(H(r))$, estimated on a finite sample (we discuss estimators for the blip in Section \ref{est_CATE}). This data set has empirical distribution $P_n$, giving each observation weight $\frac{1}{n}$. 

\subsubsection{Conditional Average Treatment Effects of Preventative CCTs in ADAPT-R}

We can intervene on the ADAPT-R's SCM to define a model on the CATE. Letting $r=1$, we define counterfactuals that quantify the initial effects of giving initial CCTs on year 1 outcomes, where if $a(1)=1$, this would set $A(1)$ equal to everyone receiving CCTs; conversely, setting $A(1)$ to $a(1)=0$ means everyone receives an intervention that is not CCTs. In this first stage, $Z(1) = \emptyset$, because patients were marginally randomized to their prevention intervention. In this way, we can define the CATE relative to ADAPT-R outcomes at time 1 as follows: $B_{P_{U,X}}(L(0)) \equiv \mathbb{E}_{P_{U,X}}[Y(1)_{a(1)=1} - Y(1)_{a(1)=0}|L(0)]$. In words, this is the average treatment effect of receiving CCTs $A(1)$ on HIV care adherence $Y(1)$, conditional on patients' baseline covariates $L(0)$.

Using our knowledge of how data were collected in the ADAPT-R study, we assume the observed data $O_i \equiv (L(0)_i, A(1)_i, L(1)_i, A(2)_i, L(2)_i) \sim P_0 \in \mathcal{M}$, $i = 1, \ldots, n$ were generated by sampling $1,815$ independent and identically distributed copies from a data-generating system contained in ADAPT-R's causal model, above. 

Further, in ADAPT-R, individuals were marginally randomized to $A(1)$, implying $Y(1)_{a(1)} \perp A(1)$ and thus $Y(1)_{a(1)} \perp A(1) |L(0)$ hold. Further, $g_{A(1),0}(A(1)=1|L(0)) = 1/3$ and $g_{A(1),0}(A(1)=0|L(0)) = 2/3$. Using this, we can identify the initial CATE as a function of ADAPT-R's observed distribution, i.e., $B_{P_{U,X}}(L(0)) = B_0(L(0))$. An estimator of the effect of receiving initial CCTs within subgroups of $L(0)$ based on ADAPT-R's empirical distribution $P_n$ is $B_n(L(0))$, an estimate of the true first-stage blip function $B_0(L(0))$.

\subsection{The Target Causal Parameter}\label{target-causal-param}

Our ultimate goal is to understand the effects of a later intervention as a function of earlier individualized effects. In particular, we will be targeting a data-adaptive parameter that examines whether the effects of later interventions are modified by a blip function estimated on previous data from the same participants in a particular sample. 

To do this, we consider an intervention on the SCM at time $s$; for $t = 0, ..., s-1, s+1,...,K$:
\begin{align*}
    L(t) &= f_{L(t)}(U_{L(t)}, \bar{L}(t-1), \bar{A}(t))\\
    A(t) &= f_{A(t)}(U_{A(t)}, \bar{Z}(t))\\
    A(s) &= a(s) \in \mathcal{A}_s \\
    L(s) &= f_{L(s)}(U_{L(s)}, \bar{L}(s-1), \bar{A}(s)),
\end{align*}
to define the true effect of $A(s)$ on $Y(s)$ conditional on the estimated CATE at time $r$, i.e.:
\begin{align}
    \mathbb{E}_{P_{U,X}}[Y(s)_{a(s)}| B_{n}(H(r))], \label{eff_causal}
\end{align}
where $r<s$.

We specify a working marginal structural model \citep{robins1999association, neugebauer2007nonparametric} $\Theta \equiv \{m_{\bm{\beta}}:\bm{\beta}\}$ for this true function of an initial blip estimate at time $r$ and a later treatment at time $s$. Our causal quantity of interest is then derived from the projection of the true effect as a function of a blip estimate onto this working model, which results in the definition $m_{\bm{\beta}_{P_{U,X}}}$ that represents this projection. 

Specifically, if $Y(s) \in [0,1]$, the following logistic working model could be used, for $a(s) \in \mathcal{A}_s$:
\begin{align}
    \logit m_{\bm{\bm{\beta}}}\left(a(s), B_{n}\left(H\left(r\right)\right)\right) & = \beta_0 + \beta_1a(s) + \beta_2B_{n}\left(H\left(r\right)\right) + \beta_3a(s)B_{n}\left(H\left(r\right)\right), \label{msm}
\end{align}
which defines the vector:
\begin{align*}
    \bm{\beta}_{P_{U,X}} =& \argmin_{\bm{\beta}} -\sum_{a(s)\in\mathcal{A}_s}\mathbb{E}_{P_{U,X}}\Big{[} h\left(a(s),B_{n}\left(H\left(r\right)\right)\right) \left( Y(s)_{a(s)}\log m_{\bm{\beta}} \left(a(s), B_{n}\left(H\left(r\right)\right)\right)\right) \\
    &+ \left(1-Y(s)_{a(s)}\right) \log \left(1- m_{\bm{\beta}}\left(a(s), B_{n}\left(H\left(r\right)\right)\right)\right)\Big{]},
\end{align*}
where $h\left(a(s),B_{n}\left(H\left(r\right)\right)\right)$ is a user-specified stabilizing weight function. For example, $h\left(a(s),B_{n}\left(H\left(r\right)\right)\right)$ could be $P(A(2) = 1 | B_n(H(r)))$ or simply set to 1. Then, $\bm{\beta}_{P_{U,X}}$ is the solution to the equation:
\begin{align*}
    0 =& \mathbb{E}_{P_{U,X}}\Bigg{[} h\left(a(s),B_{n}\left(H\left(r\right)\right)\right) \frac{\frac{d}{d\bm{\beta}_{P_{U,X}}} m_{\bm{\beta}_{P_{U,X}}}\left(a(s), B_{n}\left(H\left(r\right)\right)\right)}{m_{\bm{\beta}_{P_{U,X}}}\left(a(s), B_{n}\left(H\left(r\right)\right)\right)(1 - m_{\bm{\beta}_{P_{U,X}}}\left(a(s), B_{n}\left(H\left(r\right)\right)\right))} \times \\
    & \left( Y(s)_{a(s)}  - m_{\bm{\beta}_{P_{U,X}}}\left(a(s), B_{n}\left(H\left(r\right)\right)\right)\right) \Bigg{]},
\end{align*}
which, by the law of total expectation, can also be written as:
\begin{align*}
    0 =& \mathbb{E}_{P_{U,X}}\Bigg{[} h\left(a(s),B_{n}\left(H\left(r\right)\right)\right) \frac{\frac{d}{d\bm{\beta}_{P_{U,X}}} m_{\bm{\beta}_{P_{U,X}}}\left(a(s), B_{n}\left(H\left(r\right)\right)\right)}{m_{\bm{\beta}_{P_{U,X}}}\left(a(s), B_{n}\left(H\left(r\right)\right)\right)(1 - m_{\bm{\beta}_{P_{U,X}}}\left(a(s), B_{n}\left(H\left(r\right)\right)\right))} \times \\
    & \left( \mathbb{E}_{P_{U,X}} \left[ Y(s)_{a(s)} | H\left(s\right)\right] - m_{\bm{\beta}_{P_{U,X}}}\left(a(s), B_{n}\left(H\left(r\right)\right)\right)\right) \Bigg{]},
\end{align*}
where $H(s) \equiv (\bar{L}(s-1), \bar{A}(s-1))$. Thus, we can instead say we are interested in:
\begin{align*}
    \bm{\beta}_{P_{U,X}} =& \argmin_{\bm{\beta}} -\sum_{a(s)\in\mathcal{A}_s}\mathbb{E}_{P_{U,X}}\Big{[} h\left(a(s),B_{n}\left(H\left(r\right)\right)\right) \left( \mathbb{E}_{P_{U,X}} \left[ Y(s)_{a(s)} | H\left(s\right)\right]\log m_{\bm{\beta}} \left(a(s), B_{n}\left(H\left(r\right)\right)\right)\right) \\
    &+ \left(1-\mathbb{E}_{P_{U,X}} \left[ Y(s)_{a(s)} | H\left(s\right)\right]\right) \log \left(1- m_{\bm{\beta}}\left( a(s), B_{n}\left(H\left(r\right)\right)\right)\right)\Big{]}.
\end{align*}
In particular, our causal parameter of interest is then $\Psi^F(P_{U,X}) = \beta_3 \in \bm{\beta}_{P_{U,X}}$ from this marginal structural model, quantifying the degree of modification of the effect of $A(s)$ on $Y(s)$ by initial, individualized effects of $A(r)$ on $Y(r)$ estimated on the finite sample in hand. Here, we have kept $\mathcal{A}_s$ general such that $A(s)$ could take on more than two levels; however, in Appendix B we present a lower-dimensional marginal structural model that may be of interest if $\mathcal{A}_s \in \{0,1\}$.

\subsubsection{The Target Causal Parameter: Application to ADAPT-R}

In the context of ADAPT-R, we are interested in understanding the differential effects of discontinuing CCTs by estimated conditional effects of initiating CCTs. Exploiting ADAPT-R's sequential randomization scheme, we could additionally intervene on the model such that everyone receives CCTs, i.e., set $A(1)=1$ for all in the structural equations. In this way, we could examine effects in a world where everyone would have received initial CCTs, thereby including all participants in the evaluation of this parameter. The causal estimand is thus $\mathbb{E}_{P_{U,X}}\left[Y(2)_{a(1)=1, a(2)}| B_{n}\left(L\left(0\right)\right)\right]$, interpreted as the effect of discontinuing CCTs after receiving them, within strata of estimated initial effects.

The true causal function for $\mathbb{E}_{P_{U,X}}\left[Y(2)_{a(1)=1, a(2)}| B_{n}\left(L\left(0\right)\right)\right]$ corresponds to the counterfactual probabilities of treatment adherence under an intervention on (dis)continuation of CCTs that would have been observed for the population for each initial, individualized effect of receiving CCTs. We specify the following marginal structural working model to summarize how these counterfactual probabilities under continuation and discontinuation of CCTs vary as a function of initial, baseline covariate-specific CCT effects:
\begin{align}
    \logit m_{\bm{\bm{\beta}}}\left(a(2), B_{n}\left(L\left(0\right)\right)\right) & = \beta_0 + \beta_1a(2) + \beta_2B_{n}\left(L\left(0\right)\right) + \beta_3a(2)B_{n}\left(L\left(0\right)\right). \label{msm_adapt}
\end{align}
Letting the weights $h\left(a(2),B_{n}\left(L\left(0\right)\right)\right) = 1$ for simplicity, we can derive our target causal parameter from the projection of $\mathbb{E}_{P_{U,X}}\left[Y(2)_{a(1)=1, a(2)}| B_{n}\left(L\left(0\right)\right)\right]$ onto $m_{\bm{\beta}}\left(a(2), B_{n}\left(L\left(0\right)\right)\right)$ according to:
\begin{align*}
    \bm{\beta}_{P_{U,X}} = \argmin_{\bm{\beta}} - \sum_{a(2)\in\{0,1\}}\mathbb{E}_{P_{U,X}}\Big[ & Y(2)_{a(1)=1, a(2)} \log m_{\bm{\beta}}\left(a(2), B_{n}\left(L\left(0\right)\right)\right) \\
    & + \left(1-Y(2)_{a(1)=1, a(2)}\right) \log \left( 1 - m_{\bm{\beta}}\left(a(2), B_{n}\left(L\left(0\right)\right)\right) \right)\Big].
\end{align*}
Our causal parameter of interest is $\Psi^F(P_{U,X}) = \beta_3 \in \bm{\beta}_{P_{U,X}}$, which, in this case, quantifies how the effect of discontinuing CCTs on treatment adherence is modified by initial, individualized responses to initiating CCTs estimated on the ADAPT-R trial data. Our null hypothesis is that  $\Psi^F(P_{U,X}) = 0$, i.e., there are no differential effects of discontinuing CCTs by on individualized effects of initiating CCTs; the alternative hypothesis is $\Psi^F(P_{U,X}) \neq 0$.

\section{Identification of the Target Statistical Parameter}\label{identification}

Define the statistical target parameter of interest corresponding to the target causal parameter as a mapping $\Psi: \mathcal{M} \rightarrow \mathbb{R}$. With the point-treatment randomization $Y(s)_{a(s)} \perp A(s) | H(s)$ and positivity $g_{A(s),0}(A(s) = a(s)|H(s)) > 0 - a.e.$ conditions holding for all $a(s) \in \mathcal{A}_s$, we can identify the true effect of $A(s)$ on $Y(s)$ conditional on the estimated blip at time $r$, i.e., equation \ref{eff_causal}, as a function of the observed data distribution $P_0$; that is: $\mathbb{E}_{P_{U,X}}[Y(s)_{a(s)}| B_{n}(H(r))] = \mathbb{E}_0\left[\mathbb{E}_{0}\left[Y(s)| A(s) = a(s), H(s)\right] | B_{n}(H(r)) \right]$. As before, both in SMARTs and observational studies, some covariate/treatment combinations may be infeasible to examine, so it is imperative that relevant positivity condition holds for the causal parameter of interest. We note that this identification procedure did not require the sequential randomization and positivity assumptions, but rather the point-treatment randomization and positivity assumptions, as our causal question and thus parameter does not ask about simultaneous, joint effects of multiple $A(t)$s, but rather two interventions separately. If the causal question additionally requires intervening on other $A$ nodes (as in the ADAPT-R example; see the next section), the sequential randomization $Y(t)_{\bar{a}(t)} \perp A(t) |\bar{L}(t-1), \bar{A}(t-1) = \bar{a}(t-1)$ and positivity $g_{A(t),0}(A(t)=a(t)|\bar{L}(t-1),\bar{A}(t-1) = \bar{a}(t-1)) > 0 - a.e.$ assumptions hold in a SMART, as well.

Supposing that $Y \in [0,1]$ and the causal parameter of interest is an element of a vector of coefficients in a logistic marginal structural model, then:
\begin{align*}
    \bm{\beta}_0 =& \argmin_{\bm{\beta}} -\sum_{a(s)\in\mathcal{A}_s}\mathbb{E}_{0}\Big{[} h\left(a(s),B_{n}\left(H\left(r\right)\right)\right) \left( \mathbb{E}_0\left[Y(s)|A(s)=a(s),H(s)\right]\log m_{\bm{\beta}} \left(a(s), B_{n}\left(H\left(r\right)\right)\right)\right) \\
    &+ \left(1-\mathbb{E}_0\left[Y(s)|A(s)=a(s),H(s)\right]\right) \log \left(1- m_{\bm{\beta}}\left( a(s), B_{n}\left(H\left(r\right)\right)\right)\right)\Big{]},
\end{align*}
where $\bm{\beta}_{0}$ solves the equation:
\begin{align*}
    0 =& \mathbb{E}_{0}\Bigg{[} \sum_{a(s)\in \mathcal{A}_s}h\left(a(s), B_{n}\left(H\left(r\right)\right)\right) \frac{\frac{d}{d\bm{\beta}_{0}} m_{\bm{\beta}_{0}}\left(a(s), B_{n}\left(H\left(r\right)\right)\right)}{m_{\bm{\beta}_{0}}\left(a(s), B_{n}\left(H\left(r\right)\right)\right)(1 - m_{\bm{\beta}_{0}}\left(a(s), B_{n}\left(H\left(r\right)\right)\right))} \times \\
    & \left( \mathbb{E}_{0} \left[ Y(s)| A(s) = a(s),  H(s)\right] - m_{\bm{\beta}_{0}}\left(a(s), B_{n}\left(H\left(r\right)\right)\right)\right) \Bigg{]}.
\end{align*}
The estimation goal is now defined: using our observed data, we aim to estimate $\beta_3=\Psi(P_0)$, an element of $\bm{\beta}_{0}$.

\subsection{The Target Statistical Parameter: Application to ADAPT-R Study}

With the ADAPT-R design, it is known that randomization to Stage 2 treatment, conditional on $A(1)=1$, in ADAPT-R was based on whether or not a person had a lapse in year 1, implying $Y(2)_{a(2)} \perp A(2) | A(1)=1, Y(1)$ and thus $Y(2)_{a(2)} \perp A(2) | \bar{L}(1), A(1)=1$ hold. Additionally, it is known that: $g_{A(2),0}(A(2)=1|L(0), A(1)=1, W(1), Y(1)=0) = 1/2$; $g_{A(2),0}(A(2)=1|L(0), A(1)=1, W(1), Y(1)=1) = 1/3$; $g_{A(2),0}(A(2)=0|L(0), A(1)=1, W(1), Y(1)=1) = 2/3$; and,  $g_{A(2),0}(A(2)=0|L(0), A(1)=1, W(1), Y(1)=0) = 1/2$.

With this, we can identify the causal parameter examining the effect of discontinuing CCTs -- had everyone started them -- on later treatment adherence outcomes, conditional on the individualized effect of starting CCTs on earlier treatment adherence outcomes. This is because patients were sequentially randomized based on measured information $\bar{Z}(2)$, so the sequential randomization (i.e., $Y(2)_{\bar{a}(2)} \perp A(2) | \bar{L}(1), A(1)=1$ and $Y(1)_{\bar{a}(1)} \perp A(1) | L(0)$) and positivity (i.e., $g_{A(1),0}(A(1)=a(1)|L(0)) > 0 - a.e.$ and $g_{A(2),0}(A(2)=a(2)|\bar{L}(1), A(1)=1) > 0 - a.e.$) assumptions hold, for $a(1) \in \{0,1\}$ and $a(2) \in \{0,1\}$. We identify this longitudinal causal parameter as the following statistical parameter \citep{bangrobins2005}:
\begin{align*}
    & \mathbb{E}_{P_{U,X}}[Y(2)_{a(1)=1, a(2)} | B_{n}(L(0))] = \\
    & \mathbb{E}_0[\mathbb{E}_0[\mathbb{E}_0[Y(2)|L(0), A(1) = 1, L(1), A(2) = a(2)] | L(0), A(1)=1] | B_{n}(L(0))].
\end{align*} 
Because $Y(2)$ is binary, then:
\begin{align*}
    \bm{\beta}_0 =& \argmin_{\bm{\beta}} - \sum_{a(2)\in\{0,1\}}\mathbb{E}_{0}\Big{[} \mathbb{E}_0[\mathbb{E}_0[Y(2)|L(0), A(1) = 1, L(1), A(2) = a(2)] | L(0), A(1)=1]  \\
    &\times \log m_{\bm{\beta}} \left(a(2), B_{n}\left(L\left(0\right)\right)\right) \\
    &+ \left(1-\mathbb{E}_0[\mathbb{E}_0[Y(2)|L(0), A(1) = 1, L(1), A(2) = a(2)] | L(0), A(1)=1]\right) \\
    &\times \log \left(1- m_{\bm{\beta}}\left( a(2), B_{n}\left(L\left(0\right)\right)\right)\right)\Big{]},
\end{align*}
which solves the equation:
\begin{align*}
    0 =& \mathbb{E}_{0}\Bigg{[} \sum_{a(2)\in\{0,1\}}\frac{\frac{d}{d\bm{\beta}_{0}} m_{\bm{\beta}_{0}}\left(a(2), B_{n}\left(L\left(0\right)\right)\right)}{ m_{\bm{\beta}_{0}}\left(a(2), B_{n}\left(L\left(0\right)\right)\right)(1 -  m_{\bm{\beta}_{0}}\left(a(2), B_{n}\left(L\left(0\right)\right)\right))} \times \\
    & \left( \mathbb{E}_0[\mathbb{E}_0[Y(2)|L(0), A(1) = 1, L(1), A(2) = a(2)] | L(0), A(1)=1] - m_{\bm{\beta}_{0}}\left(a(2), B_{n}\left(L\left(0\right)\right)\right)\right) \Bigg{]}.
\end{align*}
The target statistical parameter of interest is identified as $\Psi(P_0) = \beta_3$ in $\bm{\beta}_0$, which is what we wish to estimate with the ADAPT-R data.

\section{Estimation and Inference}

This section is concerned with estimation and inference of $\Psi(P_0)$, the $\beta_3$ coefficient on the working marginal structural model that signals whether the effects of downstream interventions on downstream outcomes are modified by conditional effects of initial interventions on initial outcomes that are estimated on the sample in hand. We propose using TMLE for estimation of the parameters of the marginal structural model, paired with influence curve-based inference to generate $p$-values and confidence intervals on those estimates.

\subsection{Estimation of the CATE/Blip Function}\label{est_CATE}

Many estimators exist for the blip function (i.e., the identified CATE; e.g., see \cite{semenova2021debiased, kennedy2023towards, luedtkeSLODTR, montoya2023optimal} for reviews). For simplicity, we implement a simple plug-in/single-stage Q-learning estimator for the blip that requires estimating $Q^r_0(A(r), H(r)) \equiv \mathbb{E}_0[Y(r)|A(r), H(r)]$. Using any regression-based approach, such as SuperLearner \citep{van2007super}, one can estimate $Q^r_0$, which we call $Q^r_n$. Importantly, the chosen approach should include algorithms that allow for interactions between $A(r)$ and $H(r)$, in order to capture any initial differential effects by $H(r)$. To generate an estimate of the CATE, predict at $A(r) = 1$ and $A(r) = 0$ to generate $Q^r_n(1,H(r))$ and $Q^r_n(0,H(r))$, respectively. This provides an estimate of the CATE/blip: $B_n(H(r))=Q^r_n(1,H(r))-Q^r_n (0,H(r))$.

\subsection{Estimation and Inference of the Target Parameter}\label{est_tmle}
In Appendix C, we provide a step-by-step procedure for implementing a TMLE for $\Psi(P_0)$, which we call $\hat{\Psi}(P_n) = \beta^*_{3,n}$, drawing from \cite{petersen2014targeted} and \cite{rosenblum2010targeted}. The ltmle R package \citep{ltmlepackage, petersen2014targeted} also implements this TMLE, which we utilize for the simulations (Section \ref{sims}) and ADAPT-R data analysis (Section \ref{res_adapt}).

We now elaborate on sufficient conditions under which the aforementioned TMLE procedure produces a $\sqrt{n}$ consistent and asymptotically normal estimator for our target parameter $\Psi(P_0)$. The general theory for adaptive estimators, which we apply to our setting, was originally presented in \citet{hubbard2016}. Let $g_{A(s),n}$ and $Q^{*,s}_n$ be estimators of $g_{A(s),0}$ and $Q^s_0(A(s), H(s)) \equiv \mathbb{E}_0[Y(s)|A(s), H(s)]$, respectively. When $g_{A(s),0}$ is known (as in a SMART design), it can be correctly estimated via a parametric model; we also assume that $Q^{*,s}_n$ converges in probability to some $Q^s$ (which may differ from $Q^s_0$). We require an additional smoothness assumption, namely that the efficient influence curve for $\bm{\beta}$ (called $D^*(P)$, presented in Appendix C) when applied to an empirical distribution is a $P_0$-Donsker class and converges in quadratic mean. Under such conditions, our estimator is $\sqrt{n}$ consistent and asymptotically normal with a consistent estimator of the variance of $\bm{\beta_0}$ given by $\Sigma_n = \mathbb{E}_n[D^*(Q^{*,s}_n, g_{A(s),n}, \bm{\beta}^*_n)^2]$.  With this, we can construct asymptotically conservative 95\% confidence intervals for $\Psi(P_0)$, i.e.: $\beta^*_{3,n} \pm \Phi^{-1}(0.975)\frac{\sqrt{\Sigma_n(3,3)}}{\sqrt{n}}$. We note that one could circumvent making these smoothness assumptions by going after a sample-split specific target parameter \citep{hubbard2016} estimated with a cross-validated TMLE, although at the posssible loss of interpretability.

\section{Simulation Studies}\label{sims}

Using simulations, we evaluated the performance of the TMLE for the data-adaptive parameter $\Psi(P_0)$, the parameter representing whether there is modification of the effect of $A(s)$ on $Y(s)$ by initial estimated conditional effects $B_n(H(r))$. In particular, we investigated performance (bias, variance, mean squared error, 95\% confidence interval coverage, and power to detect an effect at the $\alpha = 0.05$ significance level) under 1,000 repetitions of sample size 1,815 (the original ADAPT-R sample size) under various generating processes: 1) a simple, two-stage SMART design intended to illustrate the differential effects this parameter could uncover (for the general case of any feasible downstream effects by initial effects -- not necessarily with the same treatment options at both stages) and 2) a more complex data generating process that mimics the ADAPT-R study in which the treatment options are the same at both stages (i.e., initiate/not initiate treatment at the first stage and, among those who initiated treatment, discontinue/continue treatment at the second stage). All simulations were implemented in R \citep{R}, and the code, simulated data, and results can be found at https://github.com/lmmontoya/effects-among-affected. Specific data generating processes for both simulations can be found in Appendix D.

\subsection{Simulation 1}

Consider two DGPs from a simple two-stage SMART design. The average effect of the initial treatment $A(1)$ on the initial outcome $Y(1)$ for both DGPs is positive: $\mathbb{E}_{P_{U,X}}[Y(1)_{a(1)=1} - Y(1)_{a(1)=0}]  \approx 0.1384$, and the blip random variable for both DGPs can take on 4 possible values, of which 3 are positive, and 1 is negative. Let $Y^1(2)$ be the second-stage outcome for DGP 1 and $Y^2(2)$ be the second-stage outcome for DGP 2. This simulation is set up such that $\mathbb{E}_{P_{U,X}}[Y^1(2)_{a(2)=1} - Y^1(2)_{a(2)=0}] = \mathbb{E}_{P_{U,X}}[Y^2(2)_{a(2)=1} - Y^2(2)_{a(2)=0}] \approx 0.1155$; i.e., the marginal second-stage risk differences are positive and the same for both data generating processes.

However, the first DGP illustrates a scenario in which the benefits of second-stage treatment accrue among those for whom the initial treatment is harmful. The risk difference conditional on a positive blip $\mathbb{E}_{P_{U,X}}[Y^1(2)_{a(2)=1} - Y^1(2)_{a(2)=0}|B_0(L(0)) > 0] \approx 0.0755$, while the risk difference conditional on a negative blip $\mathbb{E}_{P_{U,X}}[Y^1(2)_{a(2)=1} - Y^1(2)_{a(2)=0}|B_0(L(0)) < 0] = 0.2320$. The true $\beta_3$ coefficient corresponding to the working marginal structural model in equation \ref{msm} and \ref{msm_adapt} in a scenario where $B_n = B_0$ is approximately $-1.92$. We note that this value is not $\Psi(P_{U,X})$, our target causal quantity, as the latter is dependent on the finite sample at hand and thus will be different for every simulation instance.  

Conversely, the second DGP illustrates a scenario in which the benefits of second-stage treatment accrue among those for whom initial treatment is helpful. The risk difference conditional on a positive blip $\mathbb{E}_{P_{U,X}}[Y^2(2)_{a(2)=1} - Y^2(2)_{a(2)=0}|B_0(L(0)) > 0] \approx 0.1539$, while the risk difference conditional on a negative blip $\mathbb{E}_{P_{U,X}}[Y^2(2)_{a(2)=1} - Y^2(2)_{a(2)=0}|B_0(L(0)) < 0] \approx -0.0029$. Here, the true $\beta_3$ coefficient corresponding to the working marginal structural model in equation \ref{msm} and \ref{msm_adapt} in a scenario where $B_n = B_0$ is approximately $1.76$.

Thus, the first-stage and second-stage treatment effects for DGP 1 compared to DGP 2 are identical, yet the differential downstream treatment effects by initial upstream treatment effects differ between the two DGPs -- a result that was revealed by our proposed methods and would not otherwise have been uncovered by simply looking at marginal effects of initiating or discontinuing treatment, nor by examining differential effects of that treatment by each baseline or time-varying covariate in turn.

For this simulation example, we estimated $Q^2_0$, $g_{A(2),0}$ and $B_0(L(0))$ using correctly specified parametric models. Detailed performance results can be found in Table \ref{t1}. Under this estimator configuration, and with a sample size of 1,815, we saw $95.40\%$ and $95.30\%$ confidence interval coverage and were powered at $87.90\%$ and $84.50\%$ to test the null hypothesis that $\Psi(P_0) = 0$ for the DGPs corresponding to $Y^1(2)$ and $Y^2(2)$, respectively.

\begin{table}[ht]
\centering
\begin{tabular}{rrrrll}\hline& Bias & Variance & MSE & 95\% CI Cov. & Power \\\hline Simulation 1, DGP 1 & -0.0350 & 0.4862 & 0.4875 & 95.40\% & 87.90\% \\Simulation 1, DGP 2 & 0.0098 & 0.3857 & 0.3858 & 95.30\% & 84.50\% \\Simulation 2 & 0.1043 & 4.9055 & 4.9163 & 95.90\% & 87.20\% \\\hline \end{tabular}
\caption{Performance (bias, variance, mean squared error [MSE], 95\% confidence interval [CI] coverage [cov.], and power) for two simulation studies: Simulation 1 and Simulation 2. Simulation 1 illustrates two ``simple" sequentially randomized experiments (DGP 1 and 2) in which there are opposite differential downstream treatment effects by initial upstream treatment effects. Simulation 1 mimics the Adaptive Strategies for Preventing and Treating Lapses of Retention in HIV Care (ADAPT-R) trial.}
\label{t1}
\end{table}

\subsection{Simulation 2}\label{sim2}

The second simulation study mimics the ADAPT-R trial. Under this DGP, $\mathbb{E}_{P_{U,X}}[Y(1)_{a(1)=1} - Y(1)_{a(1)=0}] = 0.0554$ and $\mathbb{E}_{P_{U,X}}[Y(2)_{a(1)=1, a(2)=1} - Y(2)_{a(1)=1, a(2)=0}] = 0.0564$. As in the question asked with ADAPT-R data, here we are interested in summarizing $\mathbb{E}_{P_{U,X}}[Y(2)_{a(2), a(1)=1}|B_n(L(0))]$ using a marginal structural model, where our target causal parameter $\Psi(P_{U,X})$ is the $\beta_3$ coefficient according to equation \ref{msm_adapt}. Additionally, the true $\beta_3$ coefficient corresponding to the working marginal structural model in equation \ref{msm} and \ref{msm_adapt} in a scenario where $B_n = B_0$ is approximately $2.36$ -- meaning that there is modification of CCT discontinuation effects by CCT initiation effects. Specifically, harms of discontinuation accrue among those for whom CCT initiation is most beneficial \emph{and} benefits of discontinuation accrue among those for whom CCT initiation is least beneficial.

Here, the $Q$ factors used to estimate the marginal structural parameters were estimated using mis-specified parametric logistic regressions, while the $g$ factors were estimated using correctly specified parametric logistic regressions. The blip function was estimated with single-stage Q-learning, where the $Q$ factors were estimated using SuperLearner with a diversity of candidate algorithms ranging from the mean to a simple, parametric models to tree-based methods, with many allowing for interactions between $A(1)$ and $Y(1)$ (a detailed list of candidate algorithms included in the library can be found in Appendix D). As with the first simulation study, detailed performance results can be found in Table \ref{t1}. Under this DGP, for this estimator configuration, and with ADAPT-R's sample size of 1,815, we saw 95.90\% confidence interval coverage and were powered at 87.20\% for detecting effect modification by initial effects.

\section{Application to ADAPT-R Data}\label{res_adapt}

\subsection{Average effects}
Average effects of 1) starting CCTs and 2) stopping CCTs had everyone started were estimated using TMLE \citep{TLBBD}, where nuisance parameters were estimated using parametric models ($g$ factors were correctly specified and $Q$ factors were possibly mis-specified, where $Q$ factors were adjusted for sex and age for precision \citep{moore2009covariate}). The blip function was estimated as in Section \ref{est_CATE} and \ref{sim2} – single-stage Q-learning was implemented, where the $Q$ factors were estimated using a SuperLearner that included all baseline covariates. 

\begin{figure}[h]
    \centering
    \includegraphics[scale = .45]{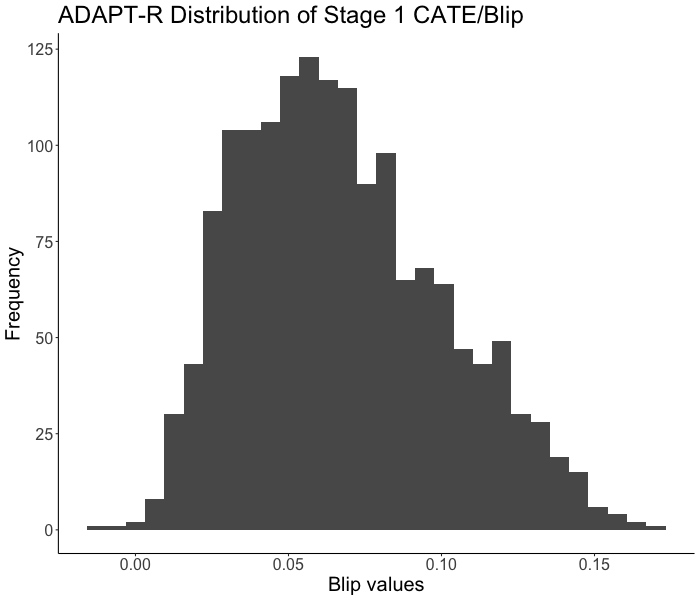}
    \caption{Distribution of conditional average treatment effect (CATE)/blip estimates of initial conditional cash transfers (CCTs) on one-year HIV care retention for the Adaptive Strategies for Preventing and Treating Lapses of Retention in HIV Care (ADAPT-R) trial.}
    \label{p1}
\end{figure}

Among the ADAPT-R sample (i.e., a sample of adults living with HIV in rural Kenya), it was shown that the probability of having no lapse in care during year 1 of the study was higher under a CCT-based treatment versus a non-CCT-based treatment (risk difference [RD]: 7.48\%, 95\% confidence interval [CI]: 3.66\%, 11.30\%). In addition, there was variability in the distribution of the average effect of receiving CCTs on remaining in care during year 1, conditional on covariates, and most of these estimates were positive (Figure \ref{p1}; range of estimated blip values: [-0.01, 0.17]). At the same time, the probability of remaining in care was significantly lower had everyone received CCTs initially and then continued versus discontinued receiving a CCT-based intervention (RD: 22.96\%, 95\% CI: 16.37\%, 29.55\%). In other words, discontinuing CCTs was harmful, on average, after initiating them. Further, the increase in benefits was significantly lower than the increase in harm (estimated difference in benefits minus harms: -15.48 \%, 95\% CI: -22.82\%, -8.13\%), suggesting that benefits of initiating CCTs do not outweigh the harms, overall.

\subsection{Effects among the affected}

Based on our results on average effects, we aimed to understand the differential effects of discontinuing CCTs by the effects of initiating CCTs -- in other words, CCT discontinuation effects among the (un)affected by initating CCTs. Specifically, we are interested in the effects on HIV care adherence. To do this, using the methods introduced above, we tested the null hypothesis that the interaction coefficient $\beta_3$ on the marginal structural model described in Equation \ref{msm_adapt} was 0. The marginal structural model was estimated using the TMLE described in Section \ref{est_tmle}, where nuisance parameters were estimated using parametric models: $g$ factors were estimated using correctly specified parametric models and $Q$ factors were estimated using possibly mis-specified parametric models, adjusting for baseline (age and sex) and time-varying covariates (time from randomization to re-reandomization) for precision \citep{moore2009covariate, montoya2023efficient}. 

We estimated our target parameter $\Psi(P_0)$ to be $\hat{\Psi}(P_n) = \beta^*_{3,n}$ = 10.71 (95 \% CI: -0.74, 22.16; $p$-value: 0.0668), implying  effect modification on the multiplicative scale at the $\alpha = 0.1$ level. If one accepts this result as a true signal (using the acceptable Type 1 error rate of $\alpha = 0.1$ often employed in evaluation of interactions \citep{selvin2004statistical}), this indicates that the effect of CCT discontinuation on HIV care adherence differs by initial responses to CCT assignment. In particular, Figure 3 illustrates that the second stage benefits of continuing CCTs were larger for those with larger initial CCT benefits compared to negligible initial CCT benefits.

\begin{figure}[h]
    \centering
    \includegraphics[scale = .6]{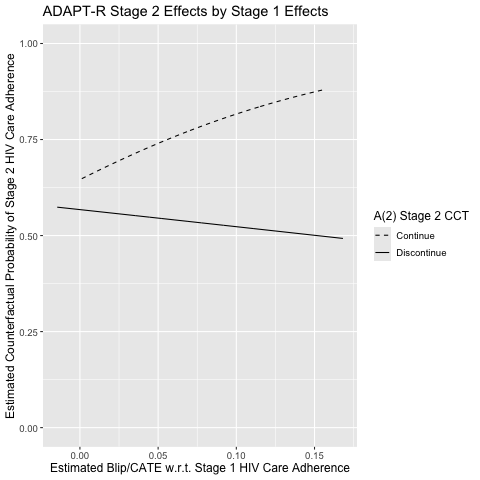}
    \caption{Differential effects of Stage 2 interventions (i.e., effect of continuing versus discontinuing conditional cash transfer [CCT]-based interventions on two-year HIV care adherence) by Stage 1 effects (i.e., conditional average effects [CATE]/blip of initial, preventative CCT on one-year HIV care adherence) for the Adaptive Strategies for Preventing and Treating Lapses of Retention in HIV Care (ADAPT-R) trial.}
    \label{p2}
\end{figure}
  
\section{Discussion}\label{discussion}

The purpose of this paper was to present a causal parameter that examines how downstream effects may be modified by initial effects of particular interventions. Specifically, we proposed a marginal structural model summarizing how the effects of a later stage treatment are modified by earlier, individualized effects; our target causal parameter was the interaction coefficient on this marginal structural model. In addition, we presented a TMLE for estimating the corresponding target statistical parameter, in addition to the efficient influence curve of this parameter, which allowed us to construct valid inference, including 95\% confidence intervals, as shown numerically via simulations. To the best of our knowledge, this approach has not yet been presented or described in the causal inference or precision medicine literature.

Importantly, this method was applied to the ADAPT-R study, a SMART intended to identify strategies for helping people living with HIV remain in their HIV care. In the primary analysis for ADAPT-R, it was shown that CCTs were an effective strategy for preventing lapses in HIV care and, overall, maintenance of CCTs was better than their discontinuation \citep{geng2023adaptive}. However, from these primary results it was not clear whether this maintenance was uniform across all participants who initially received CCTs, and further, if any (dis)continuation effects of CCTs may have been moderated by initial responses to CCT treatment. Using the methods we developed, we showed that continuation of CCTs is most effective among those who benefit from CCTs initially. This is consistent with the notion that time-limited benefits were present for a subset of participants while they were receiving CCTs, but these benefits were reversed for those participants once CCT administration stopped.

In general, the proposed methods are useful for gaining a more nuanced understanding of the effects of time-limited interventions, which could begin to inform decisions about how to deploy them. For example, knowing that CCT discontinuation is only worse among those for whom CCTs initially most helps could suggest one should continue CCTs among those for whom it initially most helps. And, among those for whom CCTs does not help initially, it does not make difference if CCTs are maintained and thus resources can be saved by not administering CCTs to the segment of the population for whom CCTs did not make much of an initial impact. Moving towards the best policy recommendation, however, could require having more decision support than just the first-stage blip (as in the marginal structural models presented in this paper). Thus, future work could extend these methods by using the initial response to treatment as input to a personalized recommendation of who should continue treatment. Concretely, this would mean using the first-stage CATE/blip of initiating the intervention as additional ``covariate" (more accurately, a dimension reduction/summary of baseline covariates) to the input of a second-stage optimal dynamic treatment rule for discontinuing the intervention.

Finally, as with all causal inference analyses, it is important to ensure that the target causal parameter one is interested in estimating is answering a relevant and feasible causal question of interest. In this case, it is imperative that the longitudinal (e.g., SMART) data in hand are generated in such a way that it can lend itself to answering the kinds of questions asked in this paper. For example, it may not be interesting or feasible to ask about differential effects of a treatment at a later stage by its earlier effects if the same treatment options do not appear in both stages of a SMART. We recommend interest in these kinds of causal questions be considered at the design stage of a study.

\section*{Acknowledgements}

Research reported in this publication was supported by NINR award R01NR018801 and NIMH award K99MH133985/R00MH133985. The content is solely the responsibility of the authors and does not necessarily represent the official views of the NIH.  \vspace*{-8pt}

\bibliographystyle{unsrt}
\bibliography{mybibilo}

\section{Appendix}

\section{Appendix A}

The following is the full list of ADAPT-R's baseline covariates $L(0)$: participant sex, age, CD4 count, depression, social support, food insecurity, treatment supporter, distance to clinic, wealth index, total work hours, total income, total discretionary spending, total fixed income, casual/wage work, work on house farm, self-employment, body mass index, pregnancy status, education level, occupation, marital status, and WHO clinical staging for HIV.

\section{Appendix B}

For a binary treatment at time $s$, one could instead define the effect among affected parameter as follow: $\mathbb{E}_{P_{U,X}}[Y(s)_{a(s)=1} - Y(s)_{a(s)=0}| B_{n}(H(r))]$. Thus, the following lower-dimensional linear working model could be used:
\begin{align*}
    \logit m_{\bm{\bm{\beta}}}\left(B_{n}\left(H\left(r\right)\right)\right) & = \beta_0 + \beta_1B_{n}\left(H\left(r\right)\right) 
\end{align*}
which defines the vector (letting stabilization weights = 1):
\begin{align*}
    \bm{\beta}_{P_{U,X}} =& \argmin_{\bm{\beta}} \mathbb{E}_{P_{U,X}} \left[\left([Y(s)_{a(s)=1} - Y(s)_{a(s)=0}] - m_{\bm{\bm{\beta}}}\left(B_{n}\left(H\left(r\right)\right)\right) \right)^2 \right].
\end{align*}
Then, $\bm{\beta}_{P_{U,X}}$ is the solution to the equation:
\begin{align*}
    0 =& \mathbb{E}_{P_{U,X}}\left\{  \frac{d}{d\bm{\beta}_{P_{U,X}}} m_{\bm{\beta}_{P_{U,X}}}\left( B_{n}\left(H\left(r\right)\right)\right)\left[(Y(s)_{a(s)=1} - Y(s)_{a(s)=0}) -  m_{\bm{\beta}_{P_{U,X}}}\left( B_{n}\left(H\left(r\right)\right)\right)\right]\right\},
\end{align*}
which can be written as:
\begin{align*}
    0 =& \mathbb{E}_{P_{U,X}}\left\{  \frac{d}{d\bm{\beta}_{P_{U,X}}} m_{\bm{\beta}_{P_{U,X}}}\left( B_{n}\left(H\left(r\right)\right)\right)\left[\mathbb{E}_{P_{U,X}}[Y(s)_{a(s)=1} - Y(s)_{a(s)=0}|H(s)] -  m_{\bm{\beta}_{P_{U,X}}}\left( B_{n}\left(H\left(r\right)\right)\right)\right]\right\},
\end{align*}
Our causal parameter of interest is then $\Psi^F(P_{U,X}) = \beta_1 \in \bm{\beta}_{P_{U,X}}$ from this marginal structural model. 

Under the point-treatment randomization $Y(s)_{a(s)} \perp A(s) | H(s)$ and positivity conditions $g_{A(s),0}(A(s) = a(s)|H(s)) > 0 - a.e.$ hold for all $a(s) \in \{0,1\}$ discussed in the main text: 
\begin{align*}
    & \mathbb{E}_{P_{U,X}}\left[Y(s)_{a(s)=1} - Y(s)_{a(s)=0}| H(s)\right]\\
    = & \mathbb{E}_{P_{U,X}}\left[Y(s)_{a(s)=1}| H(s)\right] - \mathbb{E}_{P_{U,X}}\left[Y(s)_{a(s)=0}| H(s)\right]\\
    = & \mathbb{E}_{0}\left[Y(s)| A(s) = 1, H(s)\right] - \mathbb{E}_{0}\left[Y(s)| A(s) = 0, H(s)\right].\\
\end{align*}
Then, let
\[\tilde{D}(O)=\frac{2A(s)-1}{g_{A(s)}(A(s)|\bar{L}(s-1),\bar{A}(s-1))}[Y(2)-Q^s(A(s), H(s))]+Q^s(1, H(s))-Q^s(0, H(s))\]
where $Q^s(A(s), H(s)) \equiv \mathbb{E}[Y(s)|A(s), H(s)]$. Then $\mathbb{E}_0 [\tilde{D}(O)|H(s)]=\mathbb{E}_{0}\left[Y(s)| A(s) = 1, H(s)\right] - \mathbb{E}_{0}\left[Y(s)| A(s) = 0, H(s)\right]$ if  $Q^s=Q^s_0$ or $g_{A(s)}=g_{A(s),0}$.

We can identify the true effect of $A(s)$ on $Y(s)$ conditional on the estimated blip at time $r$:
\begin{align*}
    \mathbb{E}_{P_{U,X}}[Y(s)_{a(s) = 1} - Y(s)_{a(s) = 0}| B_{n}(H(r))] &= \mathbb{E}_{P_{U,X}}\left[\mathbb{E}_{P_{U,X}}\left[Y(s)_{a(s) = 1} - Y(s)_{a(s) = 0}| H(s)\right] | B_{n}(H(r)) \right]\\
    &= \mathbb{E}_{0}\left[\mathbb{E}_0 [\tilde{D}(O)|H(s)] | B_{n}(H(r))\right]  
\end{align*}

Again, supposing that the causal parameter of interest is an element of a vector of coefficients in a linear marginal structural model, then:
\begin{align*}
    \bm{\beta}_0 =& \argmin_{\bm{\beta}} \mathbb{E}_0 \left[\left(\mathbb{E}_0 [\tilde{D}(O)|H(s)] - m_{\bm{\bm{\beta}}}\left(B_{n}\left(H\left(r\right)\right)\right) \right)^2 \right].
\end{align*}
where $\bm{\beta}_{0}$ solves the equation:
\begin{align*}
    0 =& \mathbb{E}_0\left\{  \frac{d}{d\bm{\beta}_0} m_{\bm{\beta}_0}\left( B_{n}\left(H\left(r\right)\right)\right)\left[\mathbb{E}_0 [\tilde{D}(O)|H(s)] -  m_{\bm{\beta}_0}\left( B_{n}\left(H\left(r\right)\right)\right)\right]\right\},
\end{align*}

Using our observed data, our goal would then be to estimate $\beta_1=\Psi(P_0)$, an element of $\bm{\beta}_{0}$. We note that this method is similar to the best linear predictor of the CATE concept in \citep{chernozhukov2018generic}.

\section{Appendix C}

We describe the procedure for implementing a TMLE for $\Psi(P_0)$, the target statistical parameter of interest, and constructing confidence intervals around the resulting estimate, using procedures previously described by \cite{petersen2014targeted} and \cite{ rosenblum2010targeted}.  We illustrate this under the scenario where $\mathcal{A}_s = \{0,1\}$, although this could be generalized to second-stage treatments with more than two treatment options:
\begin{enumerate}
    \item First, generate an estimator of $g_{A(s),0}$ called $g_{A(s),n}$ (e.g., a maximum likelihood estimator for $g_0$ according to a correctly specified parametric model for $g_0$, such as a logistic regression of $A(s)$ on $H(s)$). The estimated predicted probability of being treated and untreated at time $s$ is then $g_{A(s),n}(1|H(s))$ and $g_{A(s),n}(0|H(s))$, respectively.
    \item Generate an initial estimate of $Q^s_0(A(s), H(s)) \equiv \mathbb{E}_0[Y(s)|A(s), H(s)]$ by regressing $Y(s)$ on $A(s)$ and $H(s)$; call this estimator $Q^s_n$. Using this regression fit, predict at each $A(s) = 1$ and $A(s) = 0$ value to generate $Q^s_n(1,H(s))$ and $Q^s_n(0,H(s))$.
    \item Among those who received treatment $A(s)=1$, create a $n_1 \times 6$ matrix with the columns having the following vectors of length $n_1$ (where $n_1$ is the number of people who received $A(s)=1$). Call this matrix $M_1$:
    \begin{enumerate}
        \item $Y(s)$ 
        \item $A(s)$ set to 1 for all, i.e., the constant 1 repeated $n_1$ times
        \item $B_n(H(r))$ 
        \item $A(s)\times B_n(H(r))$, which in this case is the same as $B_n(H(r))$ among those with $A(s)=1$
        \item $\logit(Q^s_n(1,H(s)))$ 
        \item Weight $1/g_{A(s),n}(1|H(s))$
    \end{enumerate}
    \item Among those who did not receive treatment $A(s)=0$, create a $n_0 \times 6$ matrix with the columns having the following vectors of length $n_0$ (where $n_0$ is the number of people who received $A(s)=0$). Call this matrix $M_0$:
    \begin{enumerate}
        \item $Y(s)$ 
        \item $A(s)$ set to 0 for all, i.e., the constant 0 repeated $n_0$ times
        \item $B_n(H(r))$ 
        \item $A(s)\times B_n(H(r))$, which in this case is 0 repeated $n_0$ times 
        \item $\logit(Q^s_n(0,H(s)))$ 
        \item Weight $1/g_{A(s),n}(0|H(s))$
    \end{enumerate}
    \item Stack $M_1$ and $M_0$ together to create the $n \times 6$ matrix $M$.
    \item Using the data from the matrix $M$, fit a pooled logistic regression of $Y(s)$ on $A(s)$, $B_n(H(r))$, and $A(s)\times B_n(H(r))$ with offset $\logit(Q^s_n(A(s),H(s)))$ and weights $1/g_{A(s),n}(A(s)|H(s))$. This gives a fit $\epsilon_n = (\epsilon_{0,n}, \epsilon_{1,n}, \epsilon_{2,n}, \epsilon_{3,n})$.
    \item Generate $Q^{*,s}_n(1,H(s))$ and $Q^{*,s}_n(0,H(s))$ by using the fit from the previous step and predicting at $A(s)=1$ and $A(s)=0$, respectively. That is:
    $$Q^{*,s}_n(1,H(s)) = \logit^{-1}\left(\logit(Q^s_n(1,H(s))) + \epsilon_{0,n} + \epsilon_{1,n} + \epsilon_{2,n}B_n(H(r)) +  \epsilon_{3,n} B_n(H(r))\right)$$
    and 
    $$Q^{*,s}_n(0,H(s)) = \logit^{-1}\left(\logit(Q^s_n(0,H(s))) + \epsilon_{0,n} + \epsilon_{2,n}B_n(H(r))\right)$$
    \item Stack $Q^{*,s}_n(1,H(s))$ and $Q^{*,s}_n(0,H(s))$ to create a single vector $\bar{Q}^{*,s}_n$ of length $2\times n$. Fit a pooled logistic regression of $\bar{Q}^{*,s}_n$ on $a(s) \in \{0,1\}$ and $B_n(H(r))$ according to the model in equation 3 in the main text. This gives a TMLE of $\bm{\beta}_0$, called $\bm{\beta}^*_n$. The $\beta_3$ coefficient in $\bm{\beta}^*_n$, which we will call $\hat{\Psi}(P_n) = \beta^*_{3,n}$, gives a TMLE of the target parameter $\Psi(P_0)$.
\end{enumerate}

For a marginal structural model according to equation 3 and with $\mathcal{A}_s = \{0,1\}$ (noting, again, that this could be generalized to second-stage treatments with more than two treatment options), The efficient influence function for $\bm{\beta}$ at a distribution $P$ is $D^*(P) \equiv C^{-1}D(P)$, where $C$ is a normalizing matrix, $C = -\mathbb{E}\left[\frac{d}{d\bm{\beta}}D(P)\right]$, and
\begin{align*}
    D(P) =& \frac{h(a(s), B_n(H(r)))(Y(s) - Q^s(A(s), H(s)))}{g(A(s)|H(s))}(1, A(s), B_n(H(r)), A(s)B_n(H(r)))^T \\
    & + \sum_{a(s) \in \mathcal{A}_s}h(a(s), B_n(H(r)))(Q^s(a(s), H(s)) \\
    & - m_{\bm{\beta}} (a(s), B_{n}(H(r))))(1, a(s), B_n(H(r)), a(s)B_n(H(r)))^T.
\end{align*}

\section{Appendix D}

This section provides the exact data generating processes (DGPs) for simulation 1 and 2. 

\subsection{Simulation 1}

In Simulation 1, we simulated data according to the following two DGPs:

\begin{align*}
L(0)  = &
    \begin{cases}
      L^1(0) \sim Bernoulli(p = 0.5), \\
      L^2(0) \sim Bernoulli(p = 0.5), \\
    \end{cases}\\
A(1) \sim & Bernoulli(p = 0.5), \\
Y(1) \sim & Bernoulli(p=Q^1_0(L^1(0), L^2(0), A(1)))\\
A(2) \sim & Bernoulli(p = 0.5), \\
Y^q(2) \sim & Bernoulli(p=Q^{2,q}_0(L^1(0), L^2(0), A(1), Y(1), A(2)),
\end{align*}
where $Q^1_0(L(0), A(1))) = \logit^{-1}(L^1(0) + L^2(0) + A(1) + L^1(0)A(1) + 2L^2(0)A(1) - 5A(1)L^1(0)L^2(0))$ so that the true first-stage blip function is:
\begin{align*}
B_0(L(0))&= \logit^{-1}(L^1(0) + L^2(0) + 1 + L^1(0) + 2L^2(0) - 5L^1(0)L^2(0)) - \logit^{-1}(L^1(0) + L^2(0)).
\end{align*}
Here, $Y^q(2)$ is indexed by $q \in \{1,2\}$ such that one process generates the outcome $Y^1(2)$ according to $Q^{2,1}_0(L(0), A(1), Y(1), A(2))) = \logit^{-1}(L^1(0)A(2))$, while another generates $Y^2(2)$ according to $Q^{2,2}_0(L(0), A(1), Y(1), A(2))) = 1 - \logit^{-1}((1 - A(2))(1 - L^1(0)))$.

\subsection{Simulation 2}

This section provides the exact DGP for Simulation 2, the simulation study that mimics the ADAPT-R trial. Specifically, we simulated data according to the following DGP:

\begin{align*}
    L^1(0) &\sim Normal(\mu = 0, \sigma^2 = 1) \\
    L^2(0) &\sim Bernoulli(p = 0.5) \\
    L^3(0) &\sim Normal(\mu = L^1(0) + L^2(0), \sigma^2 = 1) \\
    A(1) &\sim Bernoulli(p = 1/3) \\ 
    W(1) &\sim Normal(\mu = L^1(0) + L^3(0) + A(1), \sigma^2 = 1) \\
    Y(1) &\sim Bernoulli(p = Q^1_0(A(1), L^1(0), L^2(0), L^3(0))) \\
    A(2) &\sim \begin{cases}
		Bernoulli(p = 1/3) & \text{if $A(1) = 1, Y(1) = 1$}\\
            Bernoulli(p = 1/2) & \text{if $A(1) = 1, Y(1) = 0$}\\
            Bernoulli(p = 1/3) & \text{if $A(1) = 0, Y(1) = 1$}\\
            Bernoulli(p = 0) & \text{if $A(1) = 0, Y(1) = 0$}\\
		 \end{cases}\\
   Y(2) &\sim Bernoulli(p = Q^2_0(L^1(0), L^2(0), L^3(0), A(1), Y(1), A(2))),
\end{align*}
where $Q^1_0(A(1), L^1(0), L^2(0), L^2(0)) = 1 - (0.5\logit^{-1}(1 - L^1(0)^2 + 3L^2(0) + 5L^3(0)^2A(1) - 4.45A(1)) + 0.5\logit^{-1}(0.5 + L^3(0) + 2L^1(0)L^2(0) + 3|L^2(0)|A(1) - 1.5A(1)))$, a function drawn and modified from \cite{luedtkeSLODTR}, and $Q^2_0(L^1(0), L^2(0), L^3(0), A(1), Y(1), A(2))) = 1 - \logit^{-1}(|L^1(0)|^{Y(1)} + Y(1) - 2A(2) + 5A(2)L^2(0) + sin(L^3(0)-4))$.

SuperLearner was used to estimate the $Q$ factors needed to estimate the blip function. The library of candidate algorithms used in the SuperLearner were: 1) \texttt{SL.mean}, the mean of the outcome, 2) three logistic regressions, each with main term for the treatment $A(1)$ and each covariate $L^j(0)$, plus an interaction $L^j(0)$ times $A(1)$, for $j \in \{1,..,3\}$, 3) a single logistic regression with all $L^j(0)$ and $A(1)$ as main terms, plus all two-way interactions between $L^j(0)$ times $A(1)$, for all $j \in \{1,..,3\}$, 4) \texttt{SL.glmnet} \citep{friedman2010regularization, simon2011regularization} , 5) \texttt{SL.bayesglm} \citep{gelman2008weakly},  6) \texttt{SL.earth} \citep{friedman1991multivariate}, and 6) \texttt{SL.randomForest} \citep{breiman2001random}. This same SuperLearner library was used for estimating the $Q$ factors for the blip in the the ADAPT-R analysis.

\end{document}